\documentclass[twocolumn,twoside,slac]{revtex4}
\usepackage{graphicx}
\usepackage{fancyhdr}
\begin{document}

\title{Clarens Client and Server Applications}


\author{Conrad D. Steenberg}
\author{Eric Aslakson, Julian J. Bunn, Harvey B. Newman,
Michael Thomas, Frank van Lingen}
\affiliation{California Institute of Technology, Pasadena, CA, 91125, USA}

\begin{abstract}

This paper describes Python, ROOT, Iguana and browser-based clients for the
Clarens web services framework. Back-end services provided include file
access, proxy escrow, virtual organization management, Storage Resource
Broker access, job execution, and a relational database analysis interface.

\end{abstract}
\maketitle
\thispagestyle{fancy}
%------------------------------------------------------------------------------
\section{Introduction\label{intro}}

The Clarens web services platform, described in a companion article,
\cite{architecture} acts as a go-between for distributed clients on the wide area
network to access services using the widely implemented XML-RPC and SOAP
data serialization standards on top of the lower-level HTTP protocol. Its
usefulness must ultimately be measured by the usefulness of the services and
clients themselves, however.

This paper is divided into two parts describing the currently implemented
services as well as clients taking advantage of these services.

\section{Services}

A rather terse overview of the current Clarens services are given below. The
reader is referred to the Clarens web page \cite{clarens_web} for full
documentation of all methods and modules.

\subsection{File access}

Accessing files on a remote machine remains the most useful service any
middleware product can provide. This is evident from the more than 40
million web servers that are deployed worldwide\cite{netcraft}. Although the
percentage of static files served is hard to gauge, the SPECweb99
\cite{specbench} benchmark puts this number at 70\% of requests.

Clarens serves files in two different ways: in response to standard HTTP GET
requests, as well as via a {\tt file.read()} service method. A virtual
server root directory can be defined for each of the above via the server
configuration file which may be any directory on the server system.

The {\tt file.read()} method takes a filename, an offset and the number of
bytes to return to the client. Error message are returned as serialized RPC
responses. Network I/O is handed off to the web server, which uses the
zero-copy {\tt sendfile()} system call where available to minimize CPU usage
and increase throughput.

Other file access methods include {\tt file.ls()} to obtain directory
listings, {\tt file.stat()} to obtain file or directory information, and
{\tt file.md5()} to obtain a hash value for checking file integrity.

%------------------------------------------------------------------------------
\subsection{Proxy escrow}

Despite the use of asymmetric key cryptography, key management remains a
problem, with private keys and certificates (credentials) having to be
present on the client system. In the case where the same credentials must be
used from different places, e.g. a person's desktop, laptop and other
computer systems, this is inconvenient, as well as degrading the security of
the credentials.

In analogy to the MyProxy \cite{myproxy} project, Clarens offers the ability
to store and retrieve short-lived, self-signed certificates (called proxy
certificates in Grid-oriented literature).

The RPC API provides the methods {\tt proxy.store()}, {\tt
proxy.retrieve()}, and {\tt proxy.delete()} for managing stored proxy
certificates. Combinations of private key/certificate pairs or
proxy/certificate pairs may be stored in this way, with the certificate
distinguished name (DN) acting as a unique identifier. The credential
information is stored in encrypted form using a symmetric cipher using a
password provided when invoking the {\tt store} method. The password itself
is not stored, for obvious reasons.

From the above it should be obvious that this approach presents a
chicken-and-egg problem of having to present credentials in order to obtain
credentials. This may be solved in two slightly different ways. Firstly, a
web portal may be constructed that takes input from the user and acts as an
intermediary to log into the Clarens server in question and retrieve the
credentials. This can be done with the portal residing on an arbitrary
machine. Secondly, the ability of Clarens to respond to HTTP GET requests
(i.e. act as as simple web server) may be used to construct such a portal on
the Clarens server itself, which has access to the server methods as an
unprivileged user. These web portals may also be accessed programmatically
from within programs or scripts.

Finally, it should be noted that it is up to the user whether to store any
credentials in this way, since it may be perceived as less secure than
keeping the credentials on possibly multiple systems' as files or in web
browsers.

%------------------------------------------------------------------------------
\subsection{Virtual organization management}

The Clarens authentication architecture \cite{architecture} is built around
the concept of a hierarchical virtual organization (VO) of groups and
subgroups of individuals identified by unique distinguished names (DNs).
These individuals may be both people or servers.

To ease administration of the VO, a set of methods is provided in the {\tt
group} module to create, delete, and list groups and their members and
administrators.

The most important of these methods are described in Table \ref{group_table}.
\begin{table}[tb]
\caption{VO management using the {\tt group} module.}\label{group_table}
\begin{tabular}{|l|p{6cm}|}
\hline
Method&Description\\
\hline
{\tt create}&Create a new group\\
{\tt delete}&Remove a group\\
{\tt add\_users}&Add members to a group\\
{\tt add\_admins}&Add administrators to a group\\
{\tt users}&Lists the group members\\
{\tt admins}&Lists the group admins\\
\hline
\end{tabular}
\end{table}

In addition to managing the VO structure, the group module also provide
methods to store, retrieve and search for certificates. Certificate
Authority certificates may similarly be searched and retrieved, but not
stored, since these certificates are used to ensure the uniqueness of client
distinguished names. Instead, CA certificates are managed by the system
administrator in the form of files stored in a designated directory.

%------------------------------------------------------------------------------
\subsection{Access control management}
The Clarens authorization architecture is built around the concept of
hierarchical access control lists for RPC methods \cite{architecture}.
To administer these ACLs, methods are provided to create and modify lists
of users and groups allowed access to a particular method, or module.

\begin{table}[tb]
\caption{ACL management using the {\tt system} module.}\label{acl_table}
\begin{tabular}{|l|p{5cm}|}
\hline
Method&Description\\
\hline
{\tt add\_acl\_allow}&Adds users and groups to the allow list of a method\\
{\tt add\_acl\_deny}&Adds users and groups to the deny list\\
{\tt add\_acl\_spec}&Create a new ACL\\
{\tt del\_acl\_spec}&Delete an ACL\\
{\tt get\_acl\_spec}&Lst ACLs \\
\hline
\end{tabular}
\end{table}

%------------------------------------------------------------------------------
\subsection{Storage Resource Broker access}

The SDSC Storage Resource Broker (SRB) provides a uniform interface to
external applications to access various storage media, including local and
remote file systems, and tape storage. It is also a client-server system,
with Clarens acting as an SRB client on behalf of its own clients.
The Clarens SRB API provides methods to initialize a connection to the SRB
server and browse, store, and retrieve files. 

This interface is not currently deployed, however, since it exposed a
critical impedance mismatch between these two client-server systems: Clarens
uses an entirely stateless connection protocol, while SRB uses a stateful
protocol. This means that in the current implementation a new SRB connection 
must be initiated upon each method invocation by the Clarens client, which
results in very poor performance.

Work is underway to remedy this mismatch by utilizing Clarens agents that
can hold persistent connections to SRB on behalf of clients. Another
approach being considered is to implement a stateful protocol interface to
Clarens itself.

%------------------------------------------------------------------------------
\subsection{Job execution}

Next to file access, the ability to execute jobs on a remote machine remains
one of the cornerstones of distributed computing. Using the Clarens
distinguished name to system user mapping \cite{architecture}, system
commands can be executed by remote users.

As is common practice, Clarens makes use of a small compiled program, called
{\tt suexec} to change the permissions of the resulting process. This
program can be more easily audited for security than the entire dependency
chain of the main Clarens code. Upon receipt of a shell command, a directory
owned by the remote user is created where the command's output and error
messages are stored. The working directory of the resultant process is also
changed to this directory, and any newly created files may be accessed
remotely using a job ID which is managed by Clarens.

It should be pointed out that this interface is not designed to schedule
jobs on Clarens servers, but is most useful for handing jobs to the
schedulers (e.g. \cite{condor}, \cite{pbs}) themselves, and retrieving the
resultant output files.

This interface is currently used to develop interactive remote analysis
using the CMS ORCA analysis package, where analysis jobs themselves become
Clarens servers (albeit less featureful ones) that can act as personal
remote application servers to allow interactivity with a long-running
analysis process on a cluster.

%------------------------------------------------------------------------------
\subsection{Relational database analysis interface}
An interface to the Stl based Object Caching And Transport System (SOCATS)
being developed by Caltech, allows remote users to query large Physics
datasets using standard relational database (SQL) queries.

SOCATS results are returned in the form of a object file formatted as a
ROOT \cite{root} tree, which can be retrieved by the above-mentioned file
access methods. A ROOT interface to Clarens is also described below, which
was successfully demonstrated at the 2002 Supercomputing conference.

%------------------------------------------------------------------------------
\section{Clients}
One of the express aims of the Clarens architecture is to use widely
deployed interfaces to lower the implementation barrier for new client
applications. The availability of HTTP, SSL and SOAP/XML-RPC 
implementations on most platforms and programmings languages helps us
achieve this goal with the minimum of new code.

\subsection{Python}
Python \cite{python} is a weakly typed, object oriented scripting language.
Its programs are compiled to a platform-independent byte-code, similar in
many ways to Java. It's built-in support for both HTTP, SSL, and XML-RPC,
combined with the rapid prototyping abilities inherent in a scripting
language made it natural to be used as the default client-side development
language.

The Python Clarens client is implemented as a pure Python class, called {\tt
clarens\_client} that takes an argument for the server URL, and optionally
the certificate and private key files to be used in the connection to its
constructor method. The constructor method initiates a connection with the
server an authenticated the user using the credentials stored in the
standard places by the Globus toolkit, including proxy credentials if they
exist, or those provided to the constructor method. 

The {\tt clarens\_client} object maps all non-local method calls to remote
procedure calls, handling serializing/deserializing of the method arguments
and return values transparently. The following example of the {\tt
echo.echo()} method demonstrates this:

\begin{verbatim}
>>> obj=clarens_client("http://server.org/")
>>> obj.echo.echo("Hello")
["Hello"]
\end{verbatim}

I.e. the {\tt echo.echo()} remote method is invoked directly from the
command line, and the result is returned as a native Python string to the
caller. The result is always returned as a list, indicated by the square
brackets.

\subsection{ROOT}

The object-oriented modular architecture of the ROOT \cite{root} analysis
package, combined with its rich set of built-in objects and wide adoption in
the high energy physics community makes it a very useful client for remote
analysis functionality.

In analogy to the Python client ,Clarens ROOT client handles authentication
transparently, and provides the user with a Clarens object to communicate
wit the remote server. This client does not do automatic
serialization/deserialization of arguments and return values to native ROOT
objects, a lower level interface must be used instead for general remote
procedure calls.

It does, however, provide a convenient interface to read remote files using
the Clarens server, via a {\tt TCWebFile} object derived from the native
{\tt TFile} object. This object provides all the functionality of the local
version, allowing transparent analysis of remote files from the ROOT command
line, scripts or compiled code by changing only the object type. Using the
interactive ROOT object browser it is possible to browse the structure and
content of remote ROOT files quickly and easily, with the ability to display
histograms and other types of plots contained in the remote files interactively
or programmatically.

The {\tt TFile} base class supports a dynamic local caching mechanism that
is used to optimize file transfers, so that the extremes of transferring the
whole remote file to the local client, as well as making a remote procedure
call for small parts of the file can be avoided, striking a balance between
bandwidth and latency constraints.

Another convenience class is the {\tt TCSystemDirectory}, derived from the
{\tt TSystemDirectory} base class. This class provides an interface to the
directory browsing functionality in the Clarens {\tt file} module, allowing
the client to interactively traverse the remote directory
structure using the ROOT object browser. Any remote ROOT files may be opened
by clicking on their icons in the browser. The  {\tt TCSystemDirectory}
class may also be used programmatically from ROOT scripts or compiled C++
programs.

\subsection{IGUANA}
IGUANA is an interactive visualization toolkit, used for detector and event
visualization, as well for interactive data analysis and presentation. It
contains support for accessing remote OpenInventor-formatted files via
Clarens for local display and manipulation. 

\subsection{Browser interface}
The web browser is currently the most widely used distributed computing
tool, bar none. Modern browsers have native support for SSL-encrypted
connections and client-side certificate authentication, making it an ideal
platform for a Clarens interface.

After initial experiments with client-side Java applets for communicating
with the server, it was decided to use the Javascript language embedded in
most browsers to handle this task. Since Clarens is able to serve web pages
in response to HTTP GET requests, the browser interface is implemented as a
series of static web pages that embed Javascript scripts to handle
communication and interface display using dynamic HTML. This implementation
eliminates the need for clients to install any additional software apart
from a web browser, which most people already have.

The browser interface uses XML-RPC for data serialization since it is by far
simpler than the more complex SOAP protocol. As with the Python interface,
argument and return value serialization/deserialization is handled
transparently by the provided Javascript libraries, made easy by the object
oriented, loosely typed nature of the language.

Functionality currently provided include browsing a remote file repository,
with the ability to download files, and virtual organization management.

%------------------------------------------------------------------------------
\section{Conclusion}
Clarens provides a growing list of services and useful client
implementations for doing distributed computing in a Grid-based environment.

\begin{acknowledgments}

This work supported by Department of Energy contract DE-FC02-01ER25459, as
part of the Particle Physics DataGrid project \cite{ppdg}, and under
National Science Foundation Grant No. 0218937.

Any opinions, findings, and conclusions or recommendations expressed in this
material are those of the authors, and do not necessarily reflect the views
of the National Science Foundation.

Clarens development is hosted by SourceForge.net
\cite{sourceforge}.\vspace{2mm}

\end{acknowledgments}

\end{document}